\documentclass[10pt,a4paper]{article}
\usepackage{amsmath}
\usepackage{amssymb}
\usepackage{amscd}
\def\h{\hbar}
\def\hA{{\hat A}}

\begin{document}

\begin{titlepage}
\title{
\hfill\parbox{4cm}
{\normalsize KUNS-1643\\{\tt hep-th/0002138}}\\
\vspace{1cm}
Noncommutative Gauge Theories\\ from Deformation Quantization}
\author{
Tsuguhiko {\sc Asakawa}\thanks{{\tt
    asakawa@gauge.scphys.kyoto-u.ac.jp}}
and
Isao {\sc Kishimoto}\thanks{{\tt
    ikishimo@gauge.scphys.kyoto-u.ac.jp}}
\\[7pt]
{\it Department of Physics, Kyoto University, Kyoto 606-8502, Japan}
}
\date{\normalsize February, 2000}
\maketitle
\thispagestyle{empty}

\begin{abstract}
\normalsize
We construct noncommutative gauge theories based on the notion of the Weyl bundle, which appears in Fedosov's construction of deformation quantization on an arbitrary symplectic manifold. These correspond to D-brane worldvolume theories in non-constant B-field and curved backgrounds in string theory. All such theories are embedded into a ``universal" gauge theory of the Weyl bundle. This shows that the combination of a background field and a noncommutative field strength has universal meaning as a field strength of the Weyl bundle. We also show that the gauge equivalence relation is a part of such a ``universal" gauge symmetry.
\end{abstract}

\end{titlepage}

\section{Introduction
\label{sec:intro} }

D-brane worldvolume theory in a constant B-field background is described by so-called noncommutative Yang-Mills theory, whose multiplicative product is the Moyal-Weyl product. This has been observed in the context of Matrix Theory in \cite{CDS}\cite{DH} and recently various research in this theory is discussed in string theory viewpoint.\footnote{
There are many papers on this subject. For example, \cite{SW}\cite{ALL} and references therein.
} However, the more general situation, i.e., in a non-constant B-field background has not been understood yet. In this paper we propose one construction of such theories from the point of view of purely worldvolume theory. The idea is as follows: the Moyal-Weyl product appears originally in the deformation quantization of the Euclidean space $\mathbb R^{2n}$. This scheme is generalized to the quantization of any symplectic or Poisson manifold and resulting product is called the star product. If we regard this not as quantized space but as noncommutative geometry, a field theory with such product is the generalization of the noncommutative Yang-Mills theory. Although this idea appears extensively in the literature \cite{HGC}, explicit construction seems not to be made.\\

The deformation quantization of a Poisson manifold $(M, \{,\})$ was first defined and investigated in \cite{Ko}.  Let $Z=C^\infty(M)[[\h]]$ be a linear space of formal power series of the deformation parameter $\h$ with coefficients in $C^\infty(M)$:
\begin{equation}
 f=\sum_{k=0}^\infty\h^kf_k
\end{equation}
Deformation quantization is an associative algebra structure on $Z$ with some associative product $*$
\begin{equation}
 f*g=\sum_{k=0}^\infty \h^k M_k(f,g)
\end{equation}
where $M_k$ are bidifferential operator such that $M_0(f,g)=fg,\ M_1(f,g)-M_1(g,f)=-i\{f,g\}$. Two star products $*_1$ and  $*_2$ are called equivalent if there exists an isomorphism of algebras $T:(Z,*_1)\rightarrow(Z,*_2)$ given by a formal power of differential operators $T=T_0+\h T_1+\cdots$. Although we regard $\h$ as some scale of noncommutativity, we use the term ``quantum" as well. \\

The problem of existence and classification up to above equivalences on an arbitrary symplectic manifold was solved by several authors \cite{BFFLS}\cite{DWL}. Finally Kontsevich \cite{Ko} solved in the case of an arbitrary Poisson manifold and relation with string theory was also suggested \cite{CF}. Above mentioned papers almost follow this line.\\

We here prefer to consider the symplectic case only, because in this case, Fedosov \cite{Fed}\cite{Fedbk} has given nice simple geometrical construction based on the Weyl algebras bundle. Let us recall his original method \cite{Fed} briefly. Since each tangent space of a symplectic manifold is a symplectic vector space, it can be quantized by usual Moyal-Weyl product. These fibers constitute a bundle of algebras, which is a sort of ``quantum tangent bundle." Then Fedosov constructed a flat connection on it, adding some quantum correction to the usual affine connection.  The flat sections of this connection can be naturally identified with $Z=C^\infty(M)[[\h]]$. So the product on fibers induces a star product on $Z$. In \S\ref{sec:Fed} we review some generalization of this method in detail.\\

In \S\ref{sec:Iso} we show that the automorphism of the Weyl algebras bundle is regarded as some (infinite dimensional) gauge transformation and this relates equivalent star products. Moreover, its subgroup, which preserve a star product, corresponds to the so-called noncommutative gauge transformation. From this observation we show that the Weyl algebra bundle itself has physical meaning as some infinite dimensional gauge bundle and noncommutative gauge field is naturally introduced by its restriction (in \S\ref{sec:ncgf}). We will construct explicitly such a gauge theory. As a result, the field strength of universal gauge field is a combination such as one in Born-Infeld action.  We also show that two such gauge fields satisfy the gauge equivalence relation. 
Further physical implications of our noncommutative gauge theories are discussed in \S\ref{sec:CD}.\\

\section{Fedosov's $*$ Product
\label{sec:Fed}}

In this section, we will recall Fedosov's construction of $*$ products on an arbitrary symplectic manifold \cite{Fed}\cite{Fedbk}. Here we deal with more general version \cite{Fedbk}, which will be necessary for our purpose.

\subsection{The Formal Weyl Algebras Bundle
\label{sec:Pre} }

We consider a symplectic manifold $(M,\Omega_0)$ of dimension $2n$ with symplectic structure $\Omega_0$ as the base space of gauge theories. First, we construct the formal Weyl algebras bundle with twisted coefficients over $M$.\\ 
Let $(L,\omega)$ be a symplectic vector bundle over $M$ of dimension $2n$, which is isomorphic to $TM$, with a fixed symplectic connection $\nabla_L$.
Denote this bundle isomorphism and its dual as
\begin{eqnarray}
\theta&:& TM\rightarrow L \nonumber\\
\delta&:& L^*\rightarrow T^*M.
\end{eqnarray}
A local symplectic frame $(e_1,\cdots,e_{2n})$ of $L$, and a dual frame $(e^1,\cdots,e^{2n})$ of $L^*$ correspond to local 1-forms $\theta^i$ on $M$ giving a basis of $T^*M$, and vector fields $X_j$ giving a dual basis of $TM$, respectively, as follows:
\begin{eqnarray}
&&\theta^i=\delta(e^i), \ e_j=\theta(X_j), \nonumber\\
&&\langle e^i,e_j\rangle_L=\langle e^i,\theta(X_j)\rangle_L = \langle\delta(e^i),X_j\rangle_{TM}=\langle\theta^i,X_j\rangle_{TM}=\delta^i_j.
\end{eqnarray}
Under the isomorphism, the symplectic form $\omega$ on $L$ is mapped to $TM$ giving a nondegenerate 2-form on $M$:
\begin{equation}
\Omega_0=-{1\over2}\omega_{ij}\theta^i\wedge\theta^j.
\end{equation}
We identify $\Omega_0$ with the symplectic form on $M$, so that it should be closed $d\Omega_0=0$. For fixed $\omega$, we will use $\theta$ to vary a symplectic structure on $M$.
We further introduce a complex vector bundle ${\cal E}$ with a connection $\nabla_{\cal E}$ and its coefficient bundle ${\rm Hom}({\cal E}, {\cal E})=:{\cal A}$ over $M$. In this paper we treat ${\cal A}$ as $U(N)$ gauge bundle for simplicity.\\

Since each fiber $L_x$ at $x\in M$ is a linear symplectic space, it can be quantized by the standard Moyal-Weyl product. The formal Weyl algebra $W_x(L,{\cal A})$ associate to $L_x$ with coefficients in ${\cal A}_x$ is defined as an noncommutative associative algebra with a unit over ${\mathbb C}$, whose elements being formal power series 
\begin{equation}
\label{eqn:element}
a(y,\h)=\sum_{2k+p\geq0,k\geq0}\h^k {1\over p!}a_{k,i_1\cdots i_p}y^{i_1}\cdots y^{i_p},
\end{equation}
where $\h$ is a formal deformation parameter, $y=(y^1,\cdots,y^{2n})$ is a linear coordinate on the fiber $L_x$ and a coefficient $a_{k,i_1\cdots i_p}\in{\cal A}_x$ is symmetric in $i_1,\cdots,i_p$. The product is defined by the Moyal-Weyl rule:
\begin{equation}
a\circ b:=\sum^\infty_{n=0}{1\over n!}\left(-{i\h\over2}\right)^n \omega^{i_1 j_1}\cdots\omega^{i_n j_n}{\partial\over\partial y^{i_1}}\cdots{\partial\over\partial y^{i_n}}a {\partial\over\partial y^{j_1}}\cdots{\partial\over\partial y^{j_n}}b,
\end{equation}
where the product of coefficients is also taken. It is easily seen that the product is associative and is independent on the choice of a basis in $L_x$. We assign degree 2 to $\h$ and degree 1 to $y^i$, so each term in (\ref{eqn:element}) has degree $2k+p\geq0$ and $\circ$ product preserves degree. \\

Taking the union $W(L,{\cal A})=\cup_{x\in M}W_x(L,{\cal A})$, we obtain the bundle of algebras over $M$, called formal Weyl algebras bundle (Weyl bundle in short.), whose sections has the form 
\begin{equation}
\label{eqn:section}
a(x,y,\h)=\sum_{2k+p\geq0,k\geq0}\h^k {1\over p!}a_{k,i_1\cdots i_p}(x)y^{i_1}\cdots y^{i_p} \,
\end{equation}
where $a_{k,i_1\cdots i_p}(x)$ is a section of ${\cal A}$, so in our case it is $N\times N$ matrix valued symmetric covariant tensor field. 
The space of sections $C^\infty(M,W(L,{\cal A}))$ also forms an associative algebra but with the {\it fiberwise} $\circ$ product. Hereafter we denote it also as  $W(L,{\cal  A})$. Note that there is a natural filtration $W(L,{\cal A})\supset W_1(L,{\cal A})\supset W_2(L,{\cal A})\supset\cdots$ with respect to the degree $2k+p$ assigned above.\\

The center $Z$ of $W(L,{\cal A})$ consists of sections which do not depend on $y^i$ and have value in multiples of the identity in ${\cal A}$ (i.e. diagonal $U(1)$ valued), thus it is naturally identified with $C^\infty(M)[[\h]]$ in \S\ref{sec:intro}. \\

A differential form on $M$ with values in $W(L,{\cal A})$ is a section of the bundle $W(L,{\cal A})\otimes{\scriptstyle \bigwedge}$ ( ${\scriptstyle \bigwedge}$ means an exterior differential algebra ${\scriptstyle \bigwedge} T^{*}M$ on $M$ ), expressed locally as
\begin{equation}
\label{eqn:aexp}
a(x,y,\h)=\sum_{2k+p\geq0,k\geq0}\h^k \sum_{q=0}^{2n}{1\over p!q!}a_{k,i_1\cdots i_p,j_1\cdots j_q}(x)y^{i_1}\cdots y^{i_p}\theta^{j_1}\wedge\cdots\wedge\theta^{j_q}.
\end{equation}
Here $a_{k,i_1\cdots i_p,j_1\cdots j_q}$ is covariant symmetric tensor in $i_1\cdots i_p$ and anti-symmetric tensor in $j_1\cdots j_q$. The sections of $W(L,{\cal A})\otimes{\scriptstyle \bigwedge}$ also form an algebra, in which the multiplication is defined by the wedge product of $\theta^j$, $\circ$ product of polynomials of $y^i$ and the matrix product of coefficients. We denote the product of two forms by the same symbol $a\circ b$ as above. Let ${\rm deg_a}a$ the rank of the differential form $a$. $W(L,{\cal A})\otimes{\scriptstyle \bigwedge}$ is formally a ${\mathbb Z}\times{\mathbb Z}$ graded algebra with respect to this $\circ$ product, this means that the $\circ$ product do not affect the degree and the ${\rm deg_a}$. Therefore the graded commutator is defined as
\begin{equation}
\label{eqn:comab}
[a,b]:=a\circ b-(-1)^{({\rm deg_a}a)({\rm deg_a}b)}b\circ a.
\end{equation}
The central $p$-form in $W(L,{\cal A})\otimes{\scriptstyle \bigwedge}$ is given in terms of (\ref{eqn:comab}) by
\begin{equation}
Z\otimes {\scriptstyle \bigwedge}^p=\{c\in W(L,{\cal A})\otimes {\scriptstyle \bigwedge}^p\ |\ [c,a]=0,\forall a\in W(L,{\cal A})\otimes {\scriptstyle \bigwedge}\},
\end{equation}
namely, it has no $y^i$ dependence and is diagonal $U(1)$ valued $p$-form.
The filtration is also satisfied: $W(L,{\cal A})\otimes{\scriptstyle \bigwedge}\supset W_1(L,{\cal A})\otimes{\scriptstyle \bigwedge}\supset W_2(L,{\cal A})\otimes{\scriptstyle \bigwedge}\supset\cdots$.\\

Next we mention a connection on the bundle $W(L,{\cal A})$. Connections $\nabla_L$ and $\nabla_{\cal E}$ generate a basic connection $\nabla=\nabla_L\otimes 1+1\otimes \nabla_{\cal E}$ on $W(L,{\cal A})$ and its induced covariant derivative $\nabla:W(L,{\cal A})\otimes{\scriptstyle \bigwedge}^q\rightarrow W(L,{\cal A})\otimes{\scriptstyle \bigwedge}^{q+1}$ may be expressed for a local symplectic frame of $L$ as follows:
\begin{eqnarray}
\label{eqn:nab}
\nabla a&=&da-\sum_{2k+p\geq1}\h^k\sum_{q=0}^{2n}{1\over(p-1)!q!}{\Gamma^m}_{i_p}\wedge a_{k,mi_1\cdots i_{p-1},j_1\cdots j_q}y^{i_1}\cdots y^{i_p}\theta^{j_1}\wedge\cdots\wedge\theta^{j_q}\nonumber\\
&&+[\Gamma_{\cal E},a]\nonumber\\
&=&da+{i\over\h}\left[{1\over2}\Gamma_{ij}y^iy^j,a\right]+[\Gamma_{\cal E},a],
\end{eqnarray}
where $d=dx^\mu\partial_\mu=\theta^iX_i$ is an exterior derivative, $\Gamma_{ij}:=\omega_{ik}{\Gamma^k}_j$ is a local connection 1-form of $\nabla_L$, 
\footnote{
         The second equal of (\ref{eqn:nab}) is valid when $\omega_{ij}=$constant (i.e., in Darboux coordinates). 
         Note $\Gamma_{ij}=\Gamma_{ji}$ when $\omega_{ij}=$constant.
         In fact, we define $\nabla_Le_i=e^j\nabla_{Lj}e_i={\Gamma^k}_{ij}e^je_k={\Gamma^k}_ie_k$, and $\nabla_L$ is symplectic connection with respect to $\omega={1\over2}\omega_{ij}e^i\wedge e^j$, therefore $0=\nabla_{Lk}\omega_{ij}:=\partial_k\omega_{ij}-\omega_{lj}{\Gamma^l}_{ik}-\omega_{il}{\Gamma^l}_{jk}=\Gamma_{jik}-\Gamma_{ijk}$ if $\partial_k\omega_{ij}=0$.
}
and $\Gamma_{\cal E}$ is a local connection 1-form of $\nabla_{\cal E}$. In the first line, ${\Gamma^i}_j$ act on tensor indices of coefficients and the term of $\Gamma_{\cal E}$ means a Lie bracket in ${\cal A}$. Both can be written in terms of the $\circ$-commutator as in the second line. This implies that $\nabla$ ($\nabla_L$) is a graded derivation with respect to $\circ$ product, namely
\begin{equation}
\nabla(a\circ b)=\nabla a\circ b+(-1)^{{\rm deg_a}a}a\circ \nabla b,\ \nabla_L(a\circ b)=\nabla_La\circ b+(-1)^{{\rm deg_a}a}a\circ \nabla_Lb.\nonumber
\end{equation}\\

There are other two canonical operators $\delta,\delta^{-1}$ on $W(L,{\cal A})\otimes{\scriptstyle \bigwedge}$, expressed locally as
\begin{eqnarray}
&&\delta=\theta^i\wedge{\partial\over\partial y^i}, \qquad \qquad \qquad \qquad \qquad \qquad  
: \ W_p(L,{\cal A})\otimes{\scriptstyle \bigwedge}^q \ \rightarrow \ W_{p-1}(L,{\cal A})\otimes{\scriptstyle \bigwedge}^{q+1} \nonumber\\
&&\delta^{-1}=\left\{
\begin{array}{ll}
y^iI(X_i){1\over p+q}  & (p+q>0)       \\
     0             & (p+q=0)
\end{array}
\right. \ \ \ : \ W_p(L,{\cal A})\otimes{\scriptstyle \bigwedge}^q \ \rightarrow \ W_{p+1}(L,{\cal A})\otimes{\scriptstyle \bigwedge}^{q-1}
\end{eqnarray}
where $I(X_i)$ is interior product, $p=$deg in $y$ and $q={\rm deg}_a$. $\delta$ is a straightforward extension of usual exterior derivative and $\delta^{-1}$ is its inverse operator (with normalization factor).
In fact, they satisfy following relations as the graded differential:
\begin{eqnarray}
\label{eqn:delrel}
       \delta^2 &=& 0,\  (\delta^{-1})^2=0 \nonumber\\
       \delta(a\circ b) &=&\delta a\circ b+(-1)^{{\rm deg_a}a} a\circ \delta b \nonumber\\
       a&=&\delta\delta ^{-1}a+\delta^{-1}\delta a+a_{00},
\end{eqnarray}
where $a_{00}$ is the $y=0$ component in $W(L,{\cal A})\otimes{\scriptstyle \bigwedge}^0$. The last relation is similar to the Hodge-de Rham decomposition.
However, we should note that $\delta$ is a purely {\it algebraic} operator containing no derivative with respect to $x$, therefore, it is also expressed as an inner derivation:
\begin{equation}
\delta a=-{i\over\h}[\omega_{ij}y^i\theta^j,a]
\end{equation}

\subsection{Abelian Connection
\label{sec:abcon}}

The main idea of Fedosov's quantization is to construct an ``Abelian" connection (defined below) on the Weyl bundle for which flat sections are identified with the quantum algebra $C^\infty(M)[[\h]]\otimes {\cal A}$. 
For this purpose, given a {\it linear} connection $\nabla$ (\ref{eqn:nab}) on $W(L,{\cal A})\otimes {\scriptstyle \bigwedge}$ in the previous section, we consider more general {\it nonlinear} connection $D$ on $W(L,{\cal A})\otimes {\scriptstyle \bigwedge}$ of the form:
\begin{equation}
\label{eqn:Dagen}
Da=\nabla a+{i\over\h}[\gamma,a],
\end{equation}
where the 1-form $\gamma$ is a {\it global} section of $W(L,{\cal A})\otimes {\scriptstyle \bigwedge}^1$. $D$ is clearly a graded derivation with respect to the $\circ$ product, i.e., $D(a\circ b)=Da\circ b+(-1)^{{\rm deg}_a a}a\circ Db$. Note that $\gamma$ in (\ref{eqn:Dagen}) is determined up to central 1-forms because it appears in the commutator
\footnote{In \cite{Fed}\cite{Fedbk} this ambiguity is fixed by the normalization condition simply setting to $0$, but we do not fix here.
}.
Simple calculation implies that
\begin{equation}
D^2a={i\over\h}[\Omega,a] \qquad \qquad  \forall a\in W(L,{\cal A})\otimes {\scriptstyle \bigwedge} .
\end{equation}
Here $\Omega$ is the curvature of $D$ given by
\begin{eqnarray}
\Omega&=&R+\nabla\gamma+{i\over\h}\gamma\circ \gamma,\nonumber\\
R&:=&{1\over2}R_{ij}y^iy^j-i\hbar R_E,\nonumber\\
R_{ij}&:=&\omega_{ik}{R^k}_j=\omega_{ik}(d{\Gamma^k}_j+{\Gamma^k}_l\wedge{\Gamma^l}_j)\nonumber\\
R_E&=&d_x\Gamma_{\cal E}+\Gamma_{\cal E}\wedge\Gamma_{\cal E},
\end{eqnarray}
where $R_{ij}$
\footnote{
        A symplectic connection $\nabla_L$ (i.e., $\nabla_L\omega_{ij}=0$) satisfies $R_{ij}=R_{ji}$.
}
 is a symplectic curvature of $\nabla_L$ and $R_E$ is a field strength of $\nabla_{\cal E}$.\\

$D$ is called an {\it Abelian} connection if $D^2a=0,\ \forall a\in W(L,{\cal A})\otimes {\scriptstyle \bigwedge}$, in other words, $\Omega$ being a central 2-form $\Omega\in Z\otimes {\scriptstyle \bigwedge}^2$. In this case Bianchi identity implies that $D\Omega=d\Omega=0$, namely $\Omega$ is closed. The condition $\Omega\in Z\otimes {\scriptstyle \bigwedge}^2$ restricts  $\gamma$.
Fedosov proved that for a given $\nabla$, there exist Abelian connections of the form
\begin{eqnarray}
\label{eqn:Abelconn}
&&Da=\nabla a-\delta a+{i\over\h}[r,a]=\nabla a+{i\over\h}[\omega_{ij}y^i\theta^j+r,a], \nonumber\\
&&\Omega=\Omega_0+\Omega_1 \nonumber\\
&&\Omega_0=-{1\over2}\omega_{ij}\theta^i\wedge\theta^j \, 
\end{eqnarray}
where $\Omega_0$ is a symplectic form on $M$ and $\Omega_1$ is a closed central 2-form, which contains at least one power of $\h$. Precisely speaking, for any choice of $\Omega=\Omega_0+{\cal O}(\h)\in Z\otimes {\scriptstyle \bigwedge}^2$ and $\mu\in W(L,{\cal A})\otimes {\scriptstyle \bigwedge}^0, {\rm deg} \mu\geq3$, the conditions that
$r$ in (\ref{eqn:Abelconn}) gives an Abelian connection are 
\begin{eqnarray}
\label{eqn:rimu}
\delta r&=&\nabla(\omega_{ij}y^i\theta^j)+R-\Omega_1+\nabla r+{i\over\h}r\circ r\nonumber\\
\delta^{-1}r&=&\mu.
\end{eqnarray}
Using the Hodge-de Rham decomposition (\ref{eqn:delrel}), these are equivalent to
\begin{equation}
\label{eqn:rita}
r=\delta\mu+\delta^{-1}\left(\nabla(\omega_{ij}y^i\theta^j)+R-\Omega_1\right)+\delta^{-1}\left(\nabla r+{i\over\h}r\circ r\right).
\end{equation}
Since $\nabla$ preserve the filtration and $\delta^{-1}$ raises it by 1, this equation can be solved {\it uniquely} by the iteration. Therefore, Abelian connections are the family parametrized by the data $\nabla$, $\Omega$ and $\mu$
\footnote{i.e., $\omega_{ij}$, $\theta^i$, $\Gamma^i_j$, $\Gamma_{\cal E}$, $\Omega_1$ and $\mu$.}.
Although it is not so desirable in the ``quantization" to exist extra parameters except for $\nabla$, there is no problem now because we regard this process as noncommutative deformations. In our case these are simply some background fields. If we further decompose $r$ to the symmetric part and others as
\begin{eqnarray}
r_s&:=&\sum_{2k+l\geq2}\h^k{1\over l!}r_{k,(i_1\cdots i_l,j)}y^{i_1}\cdots y^{i_l}\theta^j,\nonumber\\
r_a&:=&r-r_s \,
\end{eqnarray}
where $(i_1\cdots i_l,j)$ means the symmetrization of indices, then (\ref{eqn:rimu})(\ref{eqn:rita}) can be rewritten more transparently:
\begin{eqnarray}
&&\delta^{-1}r_s=\mu,\qquad  \delta r_s=0 \nonumber\\
&&\delta^{-1}r_a=0,\qquad \delta r_a=\nabla(\omega_{ij}y^i\theta^j)+R-\Omega_1+\nabla r_s+{i\over\h}r_s\circ r_s+\nabla_sr_a+{i\over\h}r_a\circ r_a,\nonumber\\
&&\rightarrow r_s = \sum_{2k+l\geq2}\h^k {1\over l!}\mu_{k,i_1\cdots i_l j}y^{i_1}\cdots y^{i_l}\theta^j, \nonumber\\
&&\rightarrow r_a =\delta^{-1}\left(\nabla(\omega_{ij}y^i\theta^j)+R-\Omega_1+\nabla r_s+{i\over\h}r_s\circ r_s\right)+\delta^{-1}\left(\nabla_s r_a+{i\over\h}r_a\circ r_a\right),\nonumber\\
\end{eqnarray}
where $\nabla_sa:=\nabla a+{i\over\h}[r_s,a]$. Easily seen from this expression, $\mu$ determines completely $r_s$, roughly speaking, which corresponds to nonlinear (quantum) corrections to $\nabla_L$.\footnote{
Note that for given $\Gamma_{ij}=\Gamma_{ijk}\theta^k$ any other symplectic connection is differ it by a completely symmetric tensor $\Delta\Gamma_{ijk}$. The leading term in $\mu$ contains this degree of freedom. The other terms are analogue of this. In \cite{EW} effects of $\mu$ are discussed in terms of a exponential map.} 
On the other hand, $r_a$ corresponds to nonlinear correction to $\Gamma_{\cal E}$ and $\delta$. Note that $\Omega_1$ appears only in the combination $i\h R_E+\Omega_1$ so that it is regarded as the correction of $U(1)$ part of $R_E$. \\

\subsection{Flat Section and $*$ product
\label{sec:flat}}

For an Abelian connection $D$, we define the space ${\scriptstyle \bigwedge} W_D$ of all {\it flat sections} in $W(L,{\cal A})\otimes{\scriptstyle \bigwedge}$ by
\begin{equation}
{{\scriptstyle \bigwedge}}W_D:=\{a\in W(L,{\cal A})\otimes {\scriptstyle \bigwedge}\  |\  Da=0\}={\rm Ker}D\cap W(L,{\cal A})\otimes{\scriptstyle \bigwedge}.
\end{equation}
Since $D$ is a graded derivation, ${\scriptstyle \bigwedge} W_D$ automatically becomes a subalgebra of $W(L,{\cal A})\otimes{\scriptstyle \bigwedge}$, namely, $a\circ b\in{\scriptstyle \bigwedge} W_D$ for $a,b\in{\scriptstyle \bigwedge} W_D$. We also define ${{\scriptstyle \bigwedge}}^pW_D$ as the space of all flat $p$-forms. Especially, for $p=0$, we denote it as $W_D:={\scriptstyle \bigwedge}^0 W_D$. Fedosov proved that $W_D$ can be naturally identified with $C^\infty(M)[[\h]]\otimes {\cal A}$, which is the quantum algebra of observables \cite{Fed}\cite{Fedbk}. In our case $C^\infty(M)[[\h]]\otimes {\cal A}$ is regarded as the space of ``fields,"  the $C^\infty(M)[[\h]]$- bimodule. In fact, let $\sigma$ denote the projection such as
\begin{eqnarray} 
 \sigma &:& W(L,{\cal A})  \rightarrow  C^\infty(M)[[\h]]\otimes {\cal A}\nonumber\\
 &   & a  \mapsto  \sigma(a):=a|_{y=0},
\end{eqnarray}
then for any $a_0\in C^\infty(M)[[\h]]\otimes {\cal A}$ there is a unique section $a\in W_D$ such that $\sigma(a)=a_0$. This is seen by rewriting the equation $Da=0$ as
\begin{equation}
\label{eqn:aita}
a=a_0+\delta^{-1}(D+\delta)a,
\end{equation}
which is solved uniquely by the iteration \cite{Fed}\cite{Fedbk}. Therefore, $\sigma$ establishes an isomorphism between $W_D$ and $C^\infty(M)[[\h]]\otimes {\cal A}$ as vector spaces. Moreover, let $Q$ be the inverse of $\sigma$  
\begin{eqnarray}
 Q &:& C^\infty(M)[[\h]]\otimes {\cal A}\rightarrow W_D\nonumber\\
 &   & a_0 \mapsto  Q(a_0)=a ,
\end{eqnarray}
then this isomorphism induces a noncommutative associative algebra structure on $C^\infty(M)[[\h]]\otimes {\cal A}$, which is a $*$ product. This is defined through $\circ$ product in $W_D$ as
\begin{equation}
\label{eqn:strdfn}
a_0*b_0=\sigma(Q(a_0)\circ Q(b_0)),\ \ a_0,b_0\in C^\infty(M)[[\h]]\otimes {\cal A}.
\end{equation}
This isomorphism will play fundamental role in later discussion in \S\ref{sec:Iso} and \S\ref{sec:ncgf}.
Note that $\circ$ product is a fiberwise product while $*$ product contains infinitely higher derivative with respect to $x$.
\\

We give a simple example of $*$ product in Appendix \ref{sec:ex}
where the noncommutativity is characterized by a constant $\vartheta^{\mu\nu}$-parameter.

\section{Isomorphism and Automorphism of Algebras
\label{sec:Iso}}

We observed in \S\ref{sec:Fed} that for each Abelian connection $D$ there exist corresponding algebras $W_D$ and $(C^\infty(M)[[\h]]\otimes {\cal A},*)$. In this section we consider two such connections $D$ and $D'$, and investigate the corresponding isomorphism between $(W_D,\circ)$ and $(W_{D'},\circ)$, which also induces an isomorphism between $(C^\infty(M)[[\h]]\otimes {\cal A},*)$ and $(C^\infty(M)[[\h]]\otimes {\cal A},*')$.\\

First, we begin with an automorphisms of the algebra $W(L,{\cal A})\otimes {\scriptstyle \bigwedge}$. For this purpose, introduce the algebra $W^+$ as an extension of $W(L,{\cal A})$ \cite{Fed}, whose elements are expressed as
\begin{equation}
U=\sum_{l=0}^\infty\sum_{2k+p=l,{\rm finite\ sum}}\h^k{1\over p!}U_{k,i_1,\cdots i_p}y^{i_1}\cdots y^{i_p},\qquad  U_{k,i_1,\cdots i_p}\in C^\infty(M)\otimes {\cal A},
\end{equation}
i.e., they may have negative powers of $\h$.
We can further introduce a group, consisting of invertible elements of $W^+$ with leading term 1 having the form
\begin{equation}
\label{eqn:Udef}
U=\exp_{\circ}\left({i\over\h}H_3\right):=\sum_{k=0}^\infty{1\over k!}\left({i\over\h}H_3\right)^k,\ H_3\in W_3'(L,{\cal A}),
\end{equation}
where $W_3'(L,{\cal A}) \ (\subset W_3(L,{\cal A}))$ consists of the elements whose diagonal $U(1)$ part are in $W_3(L,{\cal A})$ and $SU(N)$ part have at least one power of $\h$ in $W_3(L,{\cal A})$\footnote{
This restriction is required to give an automorphism of $W(L,{\cal A})\otimes {\scriptstyle \bigwedge}$ (not $W^+ \otimes {\scriptstyle \bigwedge}$). } and 
the product between $H_3$ is $\circ$ product.
It is clear from the Baker-Campbell-Hausdorff formula that such elements form a group. Now the following map
\begin{equation}
a\mapsto U^{-1}\circ a\circ U=\sum_{k=0}^\infty{1\over k!}\left({-i\over\h}\right)^k [H_3,[H_3,\cdots,[H_3,a]\cdots]]
\end{equation}
is a {\it fiberwise} automorphism of $W(L,{\cal A})\otimes {\scriptstyle \bigwedge}$. \footnote{
        This map is invariant under $U\mapsto C\circ U=CU,\ C\in Z,\exists C^{-1}$, or for a given such map $U$ is determined up to center.
}
Note that this map preserve the filtration but not the degree, namely it is a map of ``higher degree corrections" $a\mapsto a+{\cal O}(\h,y^i)$. \\

We can consider more general automorphisms of $W(L,{\cal A})\otimes {\scriptstyle \bigwedge}$ which move the supports of sections. For a diffeomorphism $f:M\rightarrow M$, its symplectic lifting to $L$ and its lifting to ${\cal E}$, we define an automorphism $A$ of the form
\begin{equation} 
A: a\mapsto f_*(U^{-1}\circ a \circ U),\ \qquad  a\in W(L,{\cal A})\otimes {\scriptstyle \bigwedge},
\end{equation}
where the liftings, the pullback $f^*$ and pushforward $f_*$ on $W(L,{\cal A})$ is defined by
\footnote{
Here $f^{0*}$ acts on $C^\infty(M)$ as the usual pullback and acts on each $y^i$ as the lifting. Since $f$ is a diffeomorphism, $f^{0*}$ on $W(L,{\cal A})\otimes {\scriptstyle \bigwedge}$ is also defined canonically. 
}
\begin{eqnarray}
\sigma_f(x)&:& L_x \rightarrow L_{f(x)},
\ \ \ {(\sigma_f(x))^k}_i\omega_{kl}(f(x)){(\sigma_f(x))^l}_j=\omega_{ij}(x),\nonumber\\
v(x)&:&{\cal E}_x \rightarrow{\cal E}_{f(x)},\nonumber\\
f^*a(x,y,\h)&:=&v(x)^{-1} a(f(x),\sigma_f(x)y,\h )v(x)=v(x)^{-1}f^{0*}a(x,y,\h)v(x),\nonumber\\
f_*a(x,y,\h)&:=&(f^*)^{-1} a=v(f^{-1}(x))a(f^{-1}(x),(\sigma_f(f^{-1}(x)))^{-1}y,\h)v(f^{-1}(x))^{-1}\nonumber\\
&=&f^0_*(v(x)a(x,y,\h)v(x)^{-1}),\nonumber\\
f^{0*}a(x,y,\h)&:=&a(f(x),\sigma_f(x)y,\h),\nonumber\\
f^0_*a(x,y,\h)&:=&a(f^{-1}(x),(\sigma_f(f^{-1}(x)))^{-1}y,\h).
\end{eqnarray}
It is easily seen if $f$ is an identity map, that $v(x)$ is a usual $U(N)$ gauge transformation and $\sigma_f(x)$ is a local $Sp(n)$ transformation, which is an analog of the local Lorentz transformation in the gravity theory (Riemannian geometry). Although a fiberwise automorphism $A$ is a quantum correction of both transformation, they play somewhat different role each other. Also in general, $v(x)$ acts only on the fiber while $\sigma_f(x)$ is a necessary part of the pullback $f^{0*}$.  Therefore, we will treat them differently as follows (the reason becomes clearer below). Since $v(x)$ acts on $W(L,{\cal A})\otimes {\scriptstyle \bigwedge}$ as the same form as $U$ does, we may include whole $v(x)$'s in the space of $U$ so that the automorphism is re-expressed as
\begin{eqnarray}
\label{eqn:auto}
&& A\ : \ a\mapsto f^0_*(U^{-1}\circ a \circ U),\ \qquad  a\in W(L,{\cal A})\otimes {\scriptstyle \bigwedge}, \nonumber \\     
&& U = \exp_{\circ}\left({i\over\h}H_2\right)=\sum_{k=0}^\infty{1\over k!}\left({i\over\h}H_2\right)^k,\ H_2\in W_2'(L,{\cal A}), 
\end{eqnarray}
Here $W_2'(L,{\cal A})$ consists of the sum of $W_3'(L,{\cal A})$ and $\h{\cal H}$ with ${\cal H}$ being a section of ${\cal A}$. We call an automorphism $A$ is a {\it ``gauge transformation"} on $W(L,{\cal A})$ if $f$ is an identity map and its symplectic lifting is trivial ${\sigma_f(x)^i}_j=\delta^i_j$, namely $f^0_*={\rm id}$. In fact $W_2'(L,{\cal A})$ forms an (infinite dimensional) Lie algebra as a linear space which includes ordinary Lie algebra $su(N)$.\\

For an Abelian connection $D$, an automorphism (\ref{eqn:auto}) defines a new connection, called an {\it image} of $D$, as usual
\begin{equation}
\label{eqn:ADA}
D'a:=AD(A^{-1}a),
\end{equation}
which is also an Abelian connection: $(D')^2 a=AD^2 (A^{-1}a)=0$.
Restricting the domain of $A$ from $W(L,{\cal A})\otimes {\scriptstyle \bigwedge}$ to ${\scriptstyle \bigwedge} W_D$, any automorphism $A$ defines an isomorphism $A\ :\ {\scriptstyle \bigwedge} W_D\rightarrow{\scriptstyle \bigwedge} W_{D'}$. In fact, if $Da=0$ then $D'a'=D'(Aa)=A(Da)=0$. Moreover, this isomorphism immediately induces an equivalence of two $*$ products. To see this, introduce a map $T:C^\infty(M)[[\h]]\otimes {\cal A}\rightarrow C^\infty(M)[[\h]]\otimes {\cal A}$ as follows
\footnote{Note that $\sigma=Q^{-1}$ on $W_D$.}:
\begin{equation}
\label{eqn:tqaq}
T:a_0\mapsto Q^{-1}A^{-1}Q'(a_0).
\end{equation}
\[
\begin{CD}
(W_D,\circ)                       @>{A}>>  (W_{D'},\circ) \\
@A{Q}AA                                       @AA{Q'}A \\
(C^\infty(M)[[\h]]\otimes {\cal A},*) @<<{T}<   (C^\infty(M)[[\h]]\otimes {\cal A},*')
\end{CD}
\]
>From the definition of $*$ products (\ref{eqn:strdfn}) we obtain
\footnote{Use following relations
\begin{eqnarray}
A^{-1}a(x,y,\h)&=&U\circ f^{0*}a(x,y,\h)\circ U^{-1},\nonumber\\
A^{-1}(a\circ b)&=&(A^{-1}a)\circ(A^{-1}b),\nonumber\\
T^{-1}&=&{Q'}^{-1}AQ,
\end{eqnarray}
}
\begin{equation}
\label{eqn:ststp}
a_0*'b_0=T^{-1}(Ta_0*Tb_0), 
\end{equation}
which shows that $T:(C^\infty(M)[[\h]]\otimes {\cal A},*')\rightarrow(C^\infty(M)[[\h]]\otimes {\cal A},*)$ is isomorphic and is nothing but the equivalence of two $*$ products
\footnote{
For detail on the equivalence of $*$ products, see \cite{BCG} for example.
}.\\

Let us see the relation between $D$ and $D'$ a little more in detail. We may rewrite an Abelian connection $D$ (\ref{eqn:Abelconn}) in a slightly different form as
\begin{eqnarray}
\label{eqn:gammaT}
Da&=&\nabla_La+{i\over\h}[\gamma_T,a],\ \ a\in W(L,{\cal A})\otimes {\scriptstyle \bigwedge}\nonumber\\
\gamma_T&:=&-i\h \Gamma_{\cal E}+\omega_{ij}y^i\theta^j+r \nonumber\\
\Omega&=&{1\over2}R_{ij}y^iy^j +\nabla_L\gamma_T+{i\over\h}\gamma_T\circ \gamma_T.
\end{eqnarray}
and in the same way for $D'$. Namely, $\Gamma_{\cal E}$ and $\gamma$ are treated together. We observe that $\gamma_T$ is nothing but the ``gauge field" with respect to the ``gauge transformations" $U$
\footnote{ Of course, $\gamma$ is the gauge field of $U$ in (\ref{eqn:Udef}) and $\Gamma_{\cal E}$ is that of $v(x)$.} in (\ref{eqn:auto}).
In fact, (\ref{eqn:ADA}) is expressed by this variables as
\begin{eqnarray}
\label{eqn:gpg}
\nabla'_L:&=&f^0_*\nabla_Lf^{0*},\nonumber\\
\gamma'_T&=&f^0_*\left(U^{-1}\circ \gamma_T\circ U-i\h U^{-1}\circ \nabla_L U+\h C_\gamma\right),
\end{eqnarray}
where the first line is adopted canonically. $C_\gamma\in Z\otimes {\scriptstyle \bigwedge}^1$ is an ambiguity of $\gamma_T$ coming from its center
\footnote{
Note that from the construction of $r$, deg $r\geq2$, so its central part has at least one power in $\h$.
}
, which is harmless in the graded commutator. In the case of a ``gauge transformation" (i.e., $f^0_*=$id. case of $A$ (\ref{eqn:auto})), $\gamma_T$ is mapped so as to be required for a ``gauge field". Further, it can be read following relations from the second line in (\ref{eqn:gpg}):
\begin{eqnarray}
\label{eqn:Uvgpg}
&&DU=U\circ {i\over\h}\left(f^{0*}\gamma'_T-\gamma_T-\h C_\gamma\right)\nonumber\\
&&f^{0*}\Omega'-\Omega-\h dC_\gamma=0\nonumber\\
&&f^{0*}\Omega_0=\Omega_0,
\end{eqnarray}
where the first line is given by simply rewriting (\ref{eqn:gpg}), the second line is obtained by operating $D$ to this first equation using $R'_{ij}y^iy^j=f^0_*(R_{ij}y^iy^j)$ and the third line is the leading term of the second equation in $\h$.
These equations mean the conditions to exist an automorphism of the form (\ref{eqn:auto}) when two arbitrary Abelian connections $D,\ D'$ are given. The last equation in (\ref{eqn:Uvgpg}) means that the map $f$ should be a symplectomorphism (symplectic diffeomorphism) on $M$ with respect to $\Omega_0$.
The second equation in (\ref{eqn:Uvgpg}) states that $\Omega$ and $f^{0*}\Omega'$ should be in the same second cohomology class. 
The first equation in (\ref{eqn:Uvgpg}) is equivalent to
\begin{equation}
U=\sigma(U)+\delta^{-1}\left((D+\delta)U-{i\over\h}U\circ \left(f^{0*}\gamma'_T-\gamma_T-\h C_\gamma\right)\right),
\end{equation}
which determines $U$ uniquely by iteration for given $\sigma(U),f^0_*,\gamma_T$ and $\gamma'_T$.
 In fact, it is proved in \cite{Fedbk} that there exists a fiberwise automorphism of the form (\ref{eqn:auto}) if the curvatures $\Omega$ and $\Omega'$ belong to the same cohomology class and their leading terms in $\h$ coincide. 
This shows that any two equivalent $*$ products or algebras $(C^\infty(M)[[\h]]\otimes {\cal A},*)$ are related by the combination of a symplectomorphism and a ``gauge transformation"
\footnote{This fact is closely related the classification problem of star products. In \cite{WX} it is proved that $[\Omega]$ is in the Hochschild cohomology of the Weyl bundle as well as in the second de-Rham cohomology of $M$.
}. \\

Although we have seen that any automorphisms $A$ of $W(L,{\cal A})\otimes{\scriptstyle \bigwedge}$ induces an isomorphism $A:{\scriptstyle \bigwedge} W_D\rightarrow{\scriptstyle \bigwedge} W_{D'}$, we are interested in a particular case that is also an automorphism of ${\scriptstyle \bigwedge} W_D$. Namely,
\begin{equation}
\label{eqn:Aiso}
Da=ADA^{-1}a,\qquad \forall a\in W(L,{\cal A})\otimes {\scriptstyle \bigwedge}.
\end{equation}
In this case $*$ product is obviously invariant, i.e., (\ref{eqn:tqaq})(\ref{eqn:ststp}) become
\begin{eqnarray}
T&=&Q^{-1}A^{-1}Q,\nonumber\\
a_0*b_0&=&T^{-1}(Ta_0*Tb_0).
\end{eqnarray}
In terms of variables in (\ref{eqn:gpg}), this condition (\ref{eqn:Aiso}) is satisfied if
\begin{eqnarray}
\label{eqn:gpgc}
\nabla_L&=&f^0_*\nabla_Lf^{0*},\nonumber\\
\gamma'_T&=&\gamma_T+\h C_\gamma',\qquad  C_\gamma'\in Z\otimes {\scriptstyle \bigwedge}^1.
\end{eqnarray}
Of course, this is a sufficient condition but not necessary. However, we here concentrate on this case, which is sufficient for our purpose. Under this restriction, possible automorphisms $A$ of ${\scriptstyle \bigwedge} W_D$ are characterized as follows: the first equation of (\ref{eqn:gpgc}) is satisfied when the symplectic lifting $\sigma_f$ is given as follows:
\begin{equation}
\label{eqn:sigf}
{\Gamma(f(x))^m}_{lk}(f^{0*}\theta^k)={\Gamma(x)^m}_{lk}\theta^k-{\left((d{\sigma_f})\sigma^{-1}_f\right)^m}_l
\end{equation}
and the the relations (\ref{eqn:Uvgpg}) become
\begin{eqnarray}
\label{eqn:Uvgpa}
&&DU=U\circ {i\over\h}\left(f^{0*}\gamma_T-\gamma_T+\h(f^{0*}C_\gamma'-C_\gamma)\right),\nonumber\\
&&\Omega'=\Omega+\h dC_\gamma',\nonumber\\
&&f^{0*}\Omega_0=\Omega_0. 
\end{eqnarray} 

If we further restrict automorphisms $A$ of $W(L,{\cal A})\otimes{\scriptstyle \bigwedge}$ to ``gauge transformations", resulting possible automorphism of ${\scriptstyle \bigwedge} W_D$ can be regarded as so-called {\it noncommutative gauge transformations}. We will explain this statement. Consider a ``gauge transformation" on $W(L,{\cal A})\otimes{\scriptstyle \bigwedge}$
\begin{equation}
\label{eqn:gauge}
A: a\mapsto U^{-1}\circ a \circ U.
\end{equation}
Since $f^0_*=$id. and the first line of (\ref{eqn:gpgc}) is automatically satisfied, the necessary condition is only the second line: $\gamma_T$ should be invariant under $U$ up to central 1-form. Conversely such $U$ is characterized by (\ref{eqn:Uvgpa}) with $f^0_*=$id.:
\begin{equation}
\label{eqn:cgam}
DU=iU(C_\gamma'-C_\gamma).
\end{equation}
Note that this is somewhat roundabout discussion because it is handled within general automorphisms. If we consider only ``gauge transformations" from the beginning, this condition is immediately derived as $D'a=D(U^{-1}\circ a\circ U)={i\over\h}[U\circ DU^{-1},a]=0$ for $Da=0$.
At any rate (\ref{eqn:cgam}) implies that $(C_\gamma'-C_\gamma)$ is closed $d(C_\gamma'-C_\gamma)=0$ because $D^2a=0$. So it is locally written as $C_\gamma'-C_\gamma=d\varphi_\gamma$ with some central function $\varphi_\gamma\in Z$, which is absorbed by the ambiguity of $U$: we may redefine $V:=U\exp(-i\varphi_\gamma)$. Then $V$ is an element of $W_D$ due to (\ref{eqn:cgam}) and (\ref{eqn:gauge}) is rewritten as
\begin{equation}
\label{eqn:gtr}
a\mapsto U^{-1}\circ a\circ U=V^{-1}\circ a\circ V  \qquad V\in W_D.
\end{equation}
This means that an ``gauge transformation'' $A$ (\ref{eqn:gauge}) preserving an Abelian connection is locally inner. Therefore, $V$ has a corresponding element in $(C^\infty(M)[[\h]]\otimes {\cal A},*)$ under the isomorphism $\sigma$ for $a\in W_D$ and (\ref{eqn:gtr}) is equivalent to
\begin{equation}
\label{eqn:gstar}
a_0\mapsto V_0^{-1}*a_0*V_0,
\end{equation}
where $a_0=\sigma(a),V_0=\sigma(V)$. This formula is the same as a usual noncommutative gauge transformation in the noncommutative Yang-Mills theory of the Moyal-Weyl type. Therefore, for an Abelian connection $D$, we call it a {\it noncommutative gauge transformation} on ${\scriptstyle \bigwedge}W_D$ (or on $(C^\infty(M)[[\h]]\otimes {\cal A},*)$ ) if a ``gauge transformation" $A$ preserves $D$. We denote it as $A_D$. Note that it is in fact local (fiberwise) transformation on $W_D$ but it is {\it not} on $(C^\infty(M)[[\h]]\otimes {\cal A},*)$ because of infinitely higher derivatives in $*$ product in the same way as the Moyal-Weyl type.\\

\section{Noncommutative Gauge Theories}
\label{sec:ncgf}

In the last section, we defined the notion of noncommutative gauge transformations on ${\scriptstyle \bigwedge}W_D$ as locally inner ``gauge transformations" of $W(L,{\cal A})\otimes{\scriptstyle \bigwedge}$. In this section we introduce an associate gauge field and construct its gauge theory. \\

\subsection{Gauge Field in the Weyl Bundle
\label{sec:gfiw}}

First we introduce a gauge field $\hA$ for ``gauge transformations" (\ref{eqn:gauge}) on $W(L,{\cal A})\otimes {\scriptstyle \bigwedge}$.\\

It has paid attention to only an Abelian connection so far: in \S\ref{sec:Fed} we constructed Abelian connections $D$ of the form (\ref{eqn:Abelconn}). In \S\ref{sec:Iso} we introduced automorphisms of the Weyl bundle $A$ (\ref{eqn:auto}) as a thing to induce isomorphisms among $W_D$'s, the spaces of flat sections with respect to $D$'s. However, in general $D$ (\ref{eqn:Dagen}) do not need to be Abelian as a connection in the Weyl bundle and $A$ (\ref{eqn:auto}) itself is defined from the first regardless of $W_D$'s. In other words, if we consider the physical theory of the Weyl bundle, we should treat $D$ (\ref{eqn:Dagen}) as dynamical variables
\footnote{Namely, $\Gamma$, $\Gamma_{\cal E}$ and $\gamma$ in (\ref{eqn:Dagen}) or $\Gamma$ and $\gamma_T$ in (\ref{eqn:gammaT}) should be done path integration.}
 and regard $A$ (\ref{eqn:auto}) as a symmetry of the system. We here concentrate on ``gauge transformations" $A$ (\ref{eqn:gauge}), therefore, the dynamical variable is $\gamma_T$ in $D$ (\ref{eqn:gammaT}) in our case. \\

As a convention, we denote such a general connection as ${\cal D}$ only to distinguish it from an Abelian connection $D$. In the same way, denote a ``gauge field" $\hA$ associated to the covariant derivative ${\cal D}:W(L,{\cal A})\otimes {\scriptstyle \bigwedge}^p\rightarrow W(L,{\cal A})\otimes {\scriptstyle \bigwedge}^{p+1}$ as:
\begin{equation}
\label{eqn:ahat}
{\cal D}a=\nabla_La-i[\hA,a],
\end{equation}
corresponding to $\gamma_T$ in $D$. $\hA$ is considered as a dynamical variable while $\nabla_L$ is fixed. The following argument is the rehash of the thing which was already done in \S\ref{sec:abcon}. As usual 
\begin{equation}
{\cal D}^2a=-i[{\hat F}_A,a],\ \ \forall a\in W(L,{\cal A})\otimes {\scriptstyle \bigwedge},
\end{equation}
where the ``field strength" ${\hat F}_A$ for $\hA$ is given by
\begin{equation}
\label{eqn:fhata}
{\hat F}_A=\nabla_L\hA-i\hA\circ\hA-{1\over2\h}R_{ij}y^iy^j.
\end{equation}
Under a ``gauge transformation" $A$ (\ref{eqn:gauge}), ${\cal D}$ is mapped to its image as
\begin{equation}
\label{eqn:caldp}
{\cal D}'a=A{\cal D}A^{-1}a,\ \ \forall a\in W(L,{\cal A})\otimes {\scriptstyle \bigwedge},
\end{equation}
This means that $\hA$ should transform as follows:
\begin{equation}
\label{eqn:ahatp}
\hA'=U^{-1}\circ \hA\circ U+iU^{-1}\circ \nabla_L U+C_A,
\end{equation}
where $C_A\in Z\otimes {\scriptstyle \bigwedge}^1$ comes from an ambiguity of the definition of $\hA$. (\ref{eqn:ahatp}) also implies
\footnote{Because we demand that ${\hat F}_A$ should be covariant under (\ref{eqn:gauge}), a condition $dC_A=0$ is imposed in (\ref{eqn:ahatp}).}
\begin{equation}
{\hat F}_A'=U^{-1}\circ {\hat F}_A \circ U.
\end{equation}
In particular, when ${\hat F}_A\in Z\otimes{\scriptstyle \bigwedge}^2$ all reduce to \S\ref{sec:abcon}, i.e., $\hA$ becomes an Abelian connection $\gamma_T$.

\subsection{Noncommutative Gauge Field
\label{sec:ncgfss}}

For a fixed ${\scriptstyle \bigwedge} W_D$, its locally inner automorphims $A_D$ has been regarded as a noncommutative gauge transformation (\S\ref{sec:Iso}). Next, we would like to consider the corresponding gauge theory. Since a noncommutative gauge transformation $A_D$ is a part of ``gauge transformation" $A$, we should introduce a noncommutative gauge filed on ${\scriptstyle \bigwedge} W_D$ by restricting a ``gauge field" $\hA$ in \S\ref{sec:gfiw} in a suitable way.\\

First, note that eq.(\ref{eqn:ahat}) can be rewritten by a simple replacement $\hA\rightarrow\hA_\gamma-{\gamma_T\over\h}$ as
\begin{eqnarray}
\label{eqn:ahatg}
{\cal D}a&=&Da-i[\hA_\gamma,a],\ \ a\in W(L,{\cal A})\otimes {\scriptstyle \bigwedge},\nonumber\\
\hA_\gamma&:=&\hA+{\gamma_T\over\h}.
\end{eqnarray}
This means that ${\cal D}$ is divided into the {\it background} $\gamma_T$, which gives Abelian connection $D$, and the {\it fluctuation} $\hA_\gamma$ around it. A choice of background $D$\footnote{
        $D$ is determined by `background' $\nabla,\mu$ and $\Omega$ (\S\ref{sec:abcon}).
} corresponding to ${\scriptstyle \bigwedge}W_D$ is changed by a ``gauge transformation" while a noncommutative gauge transformation of the present focus is a sort of background preserving one. 
Note that under a ``gauge transformation" (\ref{eqn:gauge}), eqs.(\ref{eqn:gpg})(\ref{eqn:ahatp}) imply that $\hA_\gamma$ transforms covariantly (up to center):
\begin{equation}
\label{eqn:Ag'}
\hA_\gamma'=U^{-1}\circ \hA_\gamma\circ U+C,\ \ C=C_A+C_\gamma,\ dC=0.
\end{equation}\\
In the picture above, a fixed ${\scriptstyle \bigwedge}W_D$ is interpreted as the space of fields in the corresponding noncommutative gauge theory: for example, any matter field should have $W_D$-module structure. So it is meaningful to restrict $W(L,{\cal A})\otimes {\scriptstyle \bigwedge} $ to ${\scriptstyle \bigwedge} W_D$. For an element $a\in{\scriptstyle \bigwedge} W_D$, ${\cal D}$ (\ref{eqn:ahatg}) acts on it as
\begin{equation}
\label{eqn:act}
{\cal D}a=-i[\hA_\gamma,a],\ \ a\in{\scriptstyle \bigwedge} W_D,
\end{equation}
because $Da=0$.
By the construction ${\cal D}$ is covariant under a noncommutative gauge transformation (\ref{eqn:gtr}). However, in general it is not a graded derivation of ${\scriptstyle \bigwedge} W_D$, namely ${\cal D}a$ is not necessarily an element of ${\scriptstyle \bigwedge} W_D$. Therefore, in order to define a noncommutative gauge theory we should restrict ``gauge fields" $\hA_\gamma$ such that ${\cal D}$ becomes a graded derivation of ${\scriptstyle \bigwedge} W_D$. This is the {\it definition} of a noncommutative gauge field. \\

Although it has not been mentioned until now, there is a problem in the relation between ${\scriptstyle \bigwedge} W_D$ and $C^\infty(M)[[\h]]\otimes {\cal A}$: only $W_D={\scriptstyle \bigwedge}^0 W_D$ is isomorphic to $C^\infty(M)[[\h]]\otimes {\cal A}$. In other words, $*$ product is not defined for all the differential algebra $\Omega(C^\infty(M)[[\h]]\otimes {\cal A})$. There are at least two ways which we can take: the first one is to define $*$ product for this differentioal algebra, and the second one is to treat all elements of the algebra only componentwise with respect to some fixed frame.
We here take the second way as an example and construct the theory more explicitly in \S\ref{construct}.\\

Before that we give a remark. In (\ref{eqn:act}) $\gamma_T$, contained in $\hA_\gamma$, plays the role of usual differential operator because it is written as $\nabla_La=-{i\over\h}[\gamma_T,a]$ on ${\scriptstyle \bigwedge} W_D$. $\hA_\gamma$ corresponds to a covariant coordinate discussed in \cite{MSSW} (see eq.(\ref{eqn:Ag'})).

\subsection{A Construction of Noncommutative Gauge Theory
\label{construct}}

First, we fix closed 1-forms ${\tilde \theta}^I\in Z\otimes {\scriptstyle \bigwedge}^1,\ (I=1,\cdots,2n)$, which give a basis of ${\scriptstyle \bigwedge}^1W_D$. Note that it {\it may include} any power of $\h$.
In this fixed frame ${\tilde \theta}^I$, we define the space of the fields $W_D\otimes {\tilde {\scriptstyle \bigwedge}}^p$, a subalgebra of ${\scriptstyle \bigwedge}^pW_D$, as follows:
\begin{eqnarray}
&&W_D\otimes {\tilde {\scriptstyle \bigwedge}}^p\nonumber\\
&=&\{a\in W(L,{\cal A})\otimes \wedge^p\ |\ a={1\over p!}{\tilde \theta}^{I_1}\wedge\cdots\wedge{\tilde \theta}^{I_p}Q(a_{I_1\cdots I_p}),\quad a_{I_1\cdots I_p}\in C^\infty(M)[[\h]]\otimes {\cal A}\}\nonumber\\
&\subset& {\scriptstyle \bigwedge}^pW_D,
\end{eqnarray}
where the indices of $a_{I_1\cdots I_p}$ are antisymmetric. Namely, its coefficients $Q(a_{I_1\cdots I_p})$ of any element $a$ are in $W_D$.
Therefore, we can naturally extend the isomorphism between $W_D$ and $(C^\infty(M)[[\h]]\otimes {\cal A},*)$ to an isomorphism between $W_D\otimes {\tilde {\scriptstyle \bigwedge}}^p$ and the space of $p$-forms on $(C^\infty(M)[[\h]]\otimes {\cal A},*)$ which is also given by $\sigma$ projection. 
We denote it as $(C^\infty(M)[[\h]]\otimes {\cal A}\otimes {\tilde {\scriptstyle \bigwedge}}^p,*)$.
 For example `wedge product' $\wedge$ between $p$-form $a$ and $q$-form $b$ on $(C^\infty(M)[[\h]]\otimes {\cal A},*)$ is naturally defined as follows:
\begin{eqnarray}
a\wedge b&:=&\left({1\over p!}{\tilde \theta}^{I_1}\wedge\cdots\wedge{\tilde \theta}^{I_p}\right)\wedge\left({1\over q!}{\tilde \theta}^{J_1}\wedge\cdots\wedge{\tilde \theta}^{J_q}\right)\sigma\left(Q(a_{I_1\cdots I_p})\circ Q(b_{J_1\cdots J_q})\right)\nonumber\\
&=&\left({1\over p!}{\tilde \theta}^{I_1}\wedge\cdots\wedge{\tilde \theta}^{I_p}\right)\wedge\left({1\over q!}{\tilde \theta}^{J_1}\wedge\cdots\wedge{\tilde \theta}^{J_q}\right)a_{I_1\cdots I_p}*b_{J_1\cdots J_q},
\end{eqnarray}
namely, we simply take $*$ product between coefficients of the fixed basis. 
These algebras $W_D\otimes {\tilde {\scriptstyle \bigwedge}}$ or $(C^\infty(M)[[\h]]\otimes {\cal A}\otimes {\tilde {\scriptstyle \bigwedge}},*)$ is regarded as the space of fields.
Note that since the basis ${\tilde \theta}^{I}$ may include $\h$, the algebra $(C^\infty(M)[[\h]]\otimes {\cal A}\otimes {\tilde {\scriptstyle \bigwedge}},*)$ is {\it not} a differential algebra in the usual sense. However, $(C^\infty(M)[[\h]]\otimes {\cal A},*)$ itself has no more usual meaning as the function algebra on $M$, rather the $\h$ deformed one. It is a part of $\h$ deformation of the base manifold $M$.\\

Since $W_D\otimes {\tilde {\scriptstyle \bigwedge}}\subset {\scriptstyle \bigwedge}W_D$, stricter condition for ${\cal D}$ is required so that $\hA_{\gamma}$ is a noncommutative gauge field.
In the basis ${\tilde\theta}^I$, ${\cal D}$ (\ref{eqn:act}) on $W_D\otimes {\tilde {\scriptstyle \bigwedge}}$ is written by
\begin{equation}
\label{eqn:caldag}
{\cal D}={\tilde \theta}^I{\cal D}_I,\ \ {\cal D}_Ia=-i[\hA_{\gamma I},a],\ \   a\in W_D\otimes {\tilde {\scriptstyle \bigwedge}}.
\end{equation}
${\cal D}$ becomes a graded derivation of $W_D\otimes {\tilde {\scriptstyle \bigwedge}}$ (i.e., ${\cal D}(W_D\otimes {\tilde {\scriptstyle \bigwedge}})\subset W_D\otimes {\tilde {\scriptstyle \bigwedge}}$ ) if and only if ${\cal D}_I$ is a derivation of $W_D$ (i.e., ${\cal D}_IW_D\subset W_D$).\footnote{
In general, since $D{\cal D}a=-i[D\hA_\gamma,a],\ a\in{\scriptstyle \bigwedge}W_D$, ${\cal D}$ becomes a graded derivation of ${\scriptstyle \bigwedge}W_D$ if and only if $D\hA_\gamma\in Z\otimes {\scriptstyle \bigwedge}^2$. Here a stricter condition is imposed.
}
Therefore it can be used the procedure in Appendix \ref{sec:deri} to obtain the condition that ${\cal D}_I$ becomes a derivation of $W_D$: there should exist $\Theta_I\in C^\infty(M)[[\h]]\otimes {\scriptstyle \bigwedge}^1$, such that
\begin{equation}
\label{eqn:hAcov}
D(\h \hA_{\gamma I})=\Theta_I\in C^\infty(M)[[\h]]\otimes {\scriptstyle \bigwedge}^1.
\end{equation}
Since there exists $\Phi_I$ locally such that $\Theta_I=d\Phi_I$, eq.(\ref{eqn:hAcov}) can be locally rewritten as
\begin{eqnarray}
\label{eqn:hAQP}
\h \hA_{\gamma I}&=&Q(\h \hA_{\gamma0I}-\Phi_I)+\Phi_I,\nonumber\\
\hA_{\gamma0I}&:=&\sigma(\hA_{\gamma I})=\sigma(\hA_I)+{1\over\h}\sigma(\gamma_{TI}),
\end{eqnarray}
which implies that ${\cal D}_I$ is a locally inner derivation:
\begin{equation}
\label{eqn:caldicov}
{\cal D}_Ia={i\over\h}[Q(\Phi_I),a]-i[Q(\hA_{\gamma0I}),a],\ \ a\in W_D\otimes {\tilde {\scriptstyle \bigwedge}}.
\end{equation}
This means that, under the condition above, the degrees of freedom of $\hA_{\gamma I}$ are restricted to those of $\Phi_I$ and $\hA_{\gamma0I}$. Later $\Phi_I$  together with ${\tilde\theta}^I$ becomes a sort of differential $\hat d$ with respect to the fixed background $\gamma_T$ and $\hA_{\gamma0}={\tilde \theta}^I\hA_{\gamma0I}$ is identified with a noncommutative gauge field on this background.\\

By $\sigma$ projection, (\ref{eqn:caldicov}) is reduced for $a_0=\sigma(a)\in C^\infty(M)[[\h]]\otimes {\cal A}\otimes {\tilde {\scriptstyle \bigwedge}}$ as
\begin{equation}
\label{eqn:hAcovst}
{\cal D}_*a_0={\tilde \theta}^I{\cal D}_{*I}a_0:={\tilde \theta}^I\sigma({\cal D}_IQ(a_0))={\tilde \theta}^I\left({i\over\h}[\Phi_I,a_0]_*-i[\hA_{\gamma0I},a_0]_*\right),
\end{equation}
which implies that ${\cal D}$ is a graded derivation of $(C^\infty(M)[[\h]]\otimes {\cal A}\otimes {\tilde {\scriptstyle \bigwedge}},*)$.
By operating ${\cal D}$ twice
\begin{equation}
\label{eqn:fst}
{\cal D}^2a=-{i\over2}{\tilde \theta}^I\wedge{\tilde \theta}^J\left[-i\left[Q\left({\Phi_I\over\h}-\hA_{\gamma0I}\right),Q\left({\Phi_J\over\h}-\hA_{\gamma0J}\right)\right],a\right],
\end{equation}
which is valid not only for $W_D\otimes {\tilde {\scriptstyle \bigwedge}}^p$ but also for $a\in W(L,{\cal A})\otimes {\scriptstyle \bigwedge}$.
Therefore
\begin{equation}
\label{eqn:calf}
{\hat F}_\gamma={1\over2}{\tilde \theta}^I\wedge{\tilde \theta}^J{\hat F}_{\gamma IJ}:= -{i\over2}{\tilde \theta}^I\wedge{\tilde \theta}^J\left[Q\left({\Phi_I\over\h}-\hA_{\gamma0I}\right),Q\left({\Phi_J\over\h}-\hA_{\gamma0J}\right)\right]
\end{equation}
is a field strength of $\hA_\gamma$.
In fact, ${\hat F}_\gamma$ is related to ${\hat F}_A$ (\ref{eqn:fhata}) by the relation:
\begin{equation}
\label{eqn:fof}
{\hat F}_A={\hat F}_\gamma-{1\over\h}\Omega-{1\over\h}{\tilde \theta}^I\wedge\Theta_I.
\end{equation}
${\hat F}_A$ is a field strength of the Weyl bundle and has some {\it universal} meaning: it is background independent. On the other hand, ${\hat F}_\gamma$ depends on the choice of a background $D$. By $\sigma$ projection , we get from (\ref{eqn:fst})(\ref{eqn:calf}) similar expression for $(C^\infty(M)[[\h]]\otimes {\cal A}\otimes {\tilde {\scriptstyle \bigwedge}},*)$:
\begin{eqnarray}
\label{eqn:Fstar}
{\cal D}^2a_0&=&-{i\over2}{\tilde \theta}^I\wedge{\tilde \theta}^J[{\hat F}_{\gamma*IJ},a_0]_*\nonumber\\
{\hat F}_{\gamma*}&:=&{1\over2}{\tilde \theta}^I\wedge{\tilde \theta}^J{\hat F}_{\gamma*IJ}={1\over2}{\tilde \theta}^I\wedge{\tilde \theta}^J\sigma{\hat F}_{\gamma IJ}\nonumber\\
&=&-{i\over2}{\tilde \theta}^I\wedge{\tilde \theta}^J\left[{\Phi_I\over\h}-\hA_{\gamma0I},{\Phi_J\over\h}-\hA_{\gamma0J}\right]_*.
\end{eqnarray}\\

Now we consider the degrees of freedom of ${\tilde\theta}^I$ and $\Phi_I({\rm or}\  \Theta_I)$.
In the discussion above, they are fixed by hand and are not chosen so far. We can naturally take them and interpret them as follows: There always exists a set of central functions ${\tilde \phi}^I\in Z\otimes {\scriptstyle \bigwedge}^0$ such that locally
\begin{equation}
\label{eqn:varfai}
{i\over\h}[Q({\tilde \phi}^I),Q({\tilde \phi}^J)]=-J_0^{IJ},\quad {i\over\h}[{\tilde \phi}^I,{\tilde \phi}^J]_*=-J_0^{IJ},
\end{equation}
then we choose ${\tilde \theta}^I$ and $\Theta_I$ in terms of them as
\begin{equation}
\label{eqn:bfai}
 \Phi_I=-J_{0IJ}{\tilde \phi}^J, \ \ {\tilde \theta}^I=d{\tilde \phi}^I=-J_0^{IJ}\Theta_J,
\end{equation}
where $J_0^{IJ}=-J_0^{JI}$ is constant tensor and $J_{0IJ}J_0^{JK}=\delta^K_I$. 
The second equation means that we regard ${\tilde \phi}^I$ as (quantum) coordinate functions and ${\tilde \theta}^I$ as their natural 1-form basis. One can also associate $\Phi_I$ with the dual basis ${\hat \partial}_I$. In fact, from (\ref{eqn:varfai})(\ref{eqn:bfai}) it immediately follows that
\footnote{Note that all equations in (\ref{eqn:varfai})(\ref{eqn:bfai})(\ref{eqn:strnorm}) are invariant under global $Sp(n)$ transformation.}
\begin{eqnarray}
\label{eqn:strnorm}
&{i\over\h}[Q(\Phi_I),Q(\Phi_J)]=J_{0IJ},&{i\over\h}[\Phi_I,\Phi_J]_*=J_{0IJ},\nonumber\\
&{i\over\h}[Q(\Phi_I),Q({\tilde \phi}^J)]=\delta_I^J,
&{i\over\h}[\Phi_I,{\tilde \phi}^J]_*=\delta_I^J.
\end{eqnarray}
The second line means that the first term of (\ref{eqn:caldicov}) is the partial derivative with respect to ${\tilde \phi}^I$. 
Therefore we denote it as ${\hat \partial}_I$. If we further define  
\begin{equation}
\label{eqn:dd}
{\hat d}={\tilde \theta}^I{\hat \partial}_I:={\tilde \theta}^I{i\over\h}[Q(\Phi_I),\cdot],\ \ d_*={\tilde \theta}^I\partial_{*I}:={\tilde \theta}^I{i\over\h}[\Phi_I,\cdot ]_*,
\end{equation}
then from (\ref{eqn:strnorm}) we obtain
\begin{eqnarray}
\label{eqn:hth}
&{\hat d}^2=0,&\quad\quad\quad d_*^2=0,\nonumber\\
&{\hat d}Q({\tilde \phi}^I)={\tilde \theta}^I=d{\tilde \phi}^I,&
\ d_*{\tilde \phi}^I={\tilde \theta}^I=d{\tilde \phi}^I.
\end{eqnarray}
The first line says that ${\hat d}$ and $d_*$ are differential. The second line means that the natural 1-form basis defined with respect to $d$ is consistently the natural 1-form basis defined with respect to ${\hat d}$ and $d_*$. 
In $\h\rightarrow0$ limit, ${\hat d}$ and $d_*$ reduce to usual $d$ as expected. Now the meaning of coordinate functions ${\tilde \phi}^I$ becomes clearer from the equation:
\begin{equation}
\label{eqn:om0}
\Omega_0=-{1\over2}\omega_{ij}\theta^i\wedge\theta^j
={1\over2}J_{0IJ}d{\tilde \phi}^I|_{\h=0}\wedge d{\tilde \phi}^J|_{\h=0},
\end{equation}
namely, ${\tilde \phi}^I$ are nothing but a $\h$ deformed (or quantum) Darboux coordinate, which means that ${\tilde \phi}^I=x^I+{\cal O}(\h)$ for a local Darboux coordinate $x^I$ on $M$. And as seen from (\ref{eqn:hth}), ${\tilde \theta}^I=d{\tilde \phi}^I$ are nothing but a $\h$ (or quantum) deformation of natural 1-form basis $dx^I$ of $T^* M$ and ${\hat \partial}_I$ are their deformed dual basis. These all are a realization of the deformed geometry of $M$.
\footnote{Of course this choice directly depends on the local structure of background $\nabla_L$ and $\gamma_T$. Since we are working with it locally, the differential can be identified with $\nabla_L$.}\\

In this coordinate system, eqs.(\ref{eqn:caldicov}),(\ref{eqn:hAcovst}) become 
\begin{equation}
{\cal D}a={\hat d}a-i[Q(\hA_{\gamma0}),a],\ {\cal D}_*a_0=d_*a_0-i[\hA_{\gamma0},a_0]_*,
\end{equation}
>From this expression, we may now naturally identify $Q(\hA_{\gamma0}):={\tilde \theta}^IQ(\hA_{\gamma0I})$ or $\hA_{\gamma0}:={\tilde \theta}^I\hA_{\gamma0I}$ as noncommutative gauge fields. The field strengths ${\hat F}_\gamma$ (\ref{eqn:calf}) and ${\hat F}_{\gamma*}$ (\ref{eqn:Fstar}) are also written in this basis as follows:
\begin{eqnarray}
\label{eqn:Fnorm}
{\hat F}_\gamma&=&{1\over2}{\tilde \theta}^I\wedge{\tilde \theta}^J\left({\hat \partial}_IQ(\hA_{\gamma0J})-{\hat \partial}_JQ(\hA_{\gamma0I})-i[Q(\hA_{\gamma0I}),Q(\hA_{\gamma0J})]-{J_{0IJ}\over\h}\right)
=F_\gamma-{1\over2\h}J_{0IJ}{\tilde \theta}^I\wedge{\tilde \theta}^J,\nonumber\\
{\hat F}_{\gamma*}&=&{1\over2}{\tilde \theta}^I\wedge{\tilde \theta}^J\left(\partial_{*I}\hA_{\gamma0J}-\partial_{*J}\hA_{\gamma I}-i[\hA_{\gamma0I},\hA_{\gamma0J}]_*-{J_{0IJ}\over\h}\right)
=F_{\gamma*}-{1\over2\h}J_{0IJ}{\tilde \theta}^I\wedge{\tilde \theta}^J.
\end{eqnarray}
Here $F_\gamma$ and $F_{\gamma*}$ are the field strengths of noncommutative gauge field $Q(\hA_{\gamma0})$ and $\hA_{\gamma0}$, respectively and the last constant terms come from the background. 
In this notation, eqs.(\ref{eqn:fof})(\ref{eqn:om0}) imply
\begin{equation}
\label{eqn:fagamm}
{\hat F}_A=F_\gamma-{1\over\h}\left(\Omega-\frac12 J_{0IJ}{\tilde \theta}^I\wedge{\tilde \theta}^J\right)=F_\gamma-{1\over\h}\left(\Omega_1-\frac12 J_{0IJ}{\tilde \theta}^I\wedge{\tilde \theta}^J|_{{\rm deg}\geq2}\right).
\end{equation}
This expression reminiscents $F+B+g$ appeared in Dirac-Born-Infeld action, which is familiar in low energy effective theories of strings, where $F$ is a field strength on D-brane, $B$ is NSNS 2-form field and $g$ is an induced metric on D-brane. In our case, $F_\gamma$ is a noncommutative field strength. $\Omega$ is a background 2-form, by which a $*$ product is determined. The last term corresponds to the choice of a coordinate system of $\h$ deformed geometry of $M$. This combination has some universal meaning as the field strength ${\hat F}_A$ of the Weyl bundle.\\

Under a noncommutative gauge transformation $V\in W_D$ in (\ref{eqn:Ag'}), $\hA_\gamma$ transforms covariantly as
\begin{equation}
\label{eqn:ncgtrag}
\hA'_\gamma=V^{-1}\circ \hA_\gamma\circ V+C,\ \ C=C_A+C_\gamma,\ dC=0.
\end{equation}
The central term $C$ is further restricted by consistency: 
Operating $D$ to the $I$ component of (\ref{eqn:ncgtrag}) with respect to the fixed basis ${\tilde \theta}^I$, we get using (\ref{eqn:hAcov}) that
\begin{equation}
\h D\hA'_{\gamma I}=\Theta_I+dC_I.
\end{equation}
In order to choose the same basis ${\tilde \theta}^I$ after the noncommutative gauge transformation, it is required that
\begin{equation}
dC_I=0\ \ \ \therefore C_I=C_{AI}+C_{\gamma I}={\rm constant}.
\end{equation}
Therefore, in our coordinate system (\ref{eqn:ncgtrag}) is rewritten as follows:
\begin{eqnarray}
\label{eqn:gengtr}
Q(\hA'_{\gamma0I})&=&V^{-1}\circ Q(\hA_{\gamma0I})\circ V+iV^{-1}\circ {\hat \partial}_IV+C_I\nonumber\\
&&C_I= C_{AI}+C_{\gamma I}={\rm constant},\ dC_\gamma=0,\ dC_A=0,
\end{eqnarray}
which is an expected form of the noncommutative gauge transformation. It induces the transformation of $\hA_{\gamma0I}$ on $(C^\infty(M)[[\h]]\otimes {\cal A},*)$:
\begin{equation}
\label{eqn:gengtry}
\hA'_{\gamma0I}=V_0^{-1}*\hA_{\gamma0I}*V_0+iV_0^{-1}*\partial_{*I}V_0+C_I,\ \ 
C_I={\rm constant}
\end{equation}
This is also the form of usual gauge transformation for noncommutative gauge theories up to $C$ term. $C$ term is a residual symmetry.
The gauge transformation for field strengths (\ref{eqn:Fnorm}) is
\begin{equation}
{\hat F}'_{\gamma IJ}=V^{-1}\circ {\hat F}_{\gamma IJ}\circ V,
\ \ {\hat F}'_{\gamma*IJ}=V_0^{-1}*{\hat F}_{\gamma IJ}*V_0.
\end{equation}
Then, gauge invariants on $(C^\infty(M)[[\h]]\otimes {\cal A}.*)$ can be given by the trace of them, for example:
\begin{equation}
{\rm Tr}\left({\hat F}_{\gamma*IJ}*{\hat F}_{\gamma*I'J'}J_0^{II'}J_0^{JJ'}\right),
\end{equation}
which is also global $Sp(n)$ invariants, where we denote that ${\rm Tr}$ is the trace defined in \cite{Fed}\cite{Fedbk} which contains the integration over $M$ and the usual $N\times N$ matrix trace.

\subsection{Gauge Equivalence in General
\label{sec:gegc}}

We here discuss a map from a noncommutative gauge field on $(C^\infty(M)[[\h]]\otimes {\cal A},*)$ to the one on $(C^\infty(M)[[\h]]\otimes {\cal A},*')$
which satisfies the (noncommutative) gauge equivalence relation. We call such a map as `Seiberg-Witten' map.\footnote{
	Of course, it is not quite a usual Seiberg-Witten map from an ordinary gauge field to a noncommutative gauge field.
}
In flat background (or constant $\vartheta^{\mu\nu}$ background\footnote{
        Here we define $\vartheta^{\mu\nu}$ as  $-i(x^\mu*x^\nu-x^\nu*x^\mu)$, which is usually called as a noncommutative parameter.
}), it corresponds to the one proposed in \cite{SW} in the form of the infinitesimal variation of  $\vartheta^{\mu\nu}$ at finite $\vartheta^{\mu\nu}$.
In \cite{AK}, we investigated the map required only by the gauge equivalence relation in flat backgrounds and concluded that there are ambiguities to determine a map, or some `physical input' is required to get it uniquely.
We also show the origin of these ambiguities.\\

A map between gauge fields on two different algebras $(C^\infty(M)[[\h]]\otimes {\cal A},*)$ is induced by an isomorphism between two different algebras $(W_D,\circ)$. 
A diagram of isomorphisms 
\[
\begin{CD}
W_D        @>{A}>>           W_{D'}   \\
@A{A_D}AA                      @AA{A_{D'}}A  \\
W_D        @>>{A}>           W_{D'}
\end{CD}
\]
is commutative if we take $A_{D'}=AA_DA^{-1}$. This is always possible because the diagram simply represent a group law of automorphisms in the algebra $W(L,{\cal A})$. 
This diagram means the (noncommutative) gauge equivalence relation. In fact, if we take $A_D$ given by $V\in W_D$ (noncommutative gauge transformation) and $A$ given by $U$ (``gauge transformation" for simplicity) as in \S\ref{sec:Iso}, then for an element of $W_{D'}$
\begin{equation}
A_{D'}a:=AA_DA^{-1}a
=(U^{-1}\circ V\circ U)^{-1}\circ a \circ (U^{-1}\circ V\circ U),
\end{equation}
so that $A_{D'}$ is a noncommutative gauge transformation on $W_{D'}$ given by $V'=U^{-1}\circ V \circ U\in W_{D'}$. 
Here the transformation by $U$ corresponds to the change of $\vartheta^{\mu\nu}$ in \cite{SW}\cite{AK} because $*$ product varies by $U$
\footnote{
Infinitesimal transformation $A$ corresponds to $\delta\theta$ in \cite{AK}.
}.
The automorphism $A_{D'}$ is different from $A_D$
and the map from $V$ to $V'$ corresponds to a `Seiberg-Witten' map of noncommutative gauge parameters in \cite{SW}\cite{AK}.\footnote{
$A_D$ and $A_{D'}$ correspond to ${\hat \delta}_{\hat \lambda}$ and ${\tilde \delta}_{\tilde \lambda}$ in \cite{AK}.
} 
If we apply above isomorphisms twice (denote as $U_1$ and $U_2$), then the noncommutativity $U_1\circ U_2 \neq U_2\circ U_1$ of the ``gauge transformation" (or simply that of $\circ$ product) produces the ``path dependence" investigated in \cite{AK}.\\

>From the diagram we can read a `Seiberg-Witten' map \cite{SW}\cite{AK} between noncommutative gauge fields on $(C^\infty(M)[[\h]]\otimes {\cal A},*)$ and $(C^\infty(M)[[\h]]\otimes {\cal A},*')$ by using (\ref{eqn:Ag'})(\ref{eqn:hAQP}):
\begin{eqnarray}
\label{eqn:SWmap}
{\tilde \theta}^{'I'}\hA'_{\gamma0I'}
&=&{\tilde \theta}^I\sigma\left(U^{-1}\circ Q(\hA_{\gamma0I})\circ U\right)
-{1\over\h}{\tilde \theta}^I\sigma\left(U^{-1}\circ Q(\Phi_I)\circ U\right)
+{1\over\h}{\tilde \theta}^I\Phi_I+C,\nonumber\\
&&dC=0,
\end{eqnarray}
where ${\tilde \theta}^I$ and ${\tilde \theta}^{'I'}$ are different frames because backgrounds are not the same.

In this case, a ``gauge transformation" is nothing but a `Seiberg-Witten' map which satisfies the (noncommutative) gauge equivalence relation,\footnote{
          This interpretation is closely related to recent works \cite{Cor}\cite{Oku}\cite{CHL}\cite{JS}.
} and its ambiguities come from the noncommutativity of two ``gauge transformations." In the same way, we can include  $f^0_*$ to the above $A$ i.e., consider general automorphism (\ref{eqn:auto}), but in that case the map between ${\hat A}_{\gamma}$ and ${\hat A'}_{\gamma}$ becomes more complicated.

\section{Conclusion and Discussion
\label{sec:CD}}

We have constructed noncommutative gauge theories on an arbitrary symplectic manifold $M$ in rather general situation. To obtain a noncommutative associative algebra $(C^\infty(M)[[\h]]\otimes {\cal A},*)$, Fedosov's construction \cite{Fedbk} of deformation quantization was used. In \S\ref{sec:Fed} we first introduced the notion of the Weyl bundle $W(L,{\cal A})$, an Abelian connections $D$, and an algebra of flat sections $W_D$, which is isomorphic to the algebra $(C^\infty(M)[[\h]]\otimes {\cal A},*)$. Then in \S\ref{sec:Iso} we discussed automorphisms of the Weyl bundle and induced isomorphisms among $W_D$'s. In \S\ref{sec:ncgf} we introduced a gauge field $\hA$ associated with ``gauge transformations" on the Weyl bundle and obtained a noncommutative gauge filed ${\hA}_\gamma$ by the suitable restriction on $W_D$. This gives corresponding noncommutative gauge field $\hA_{\gamma0}$ on $(C^\infty(M)[[\h]]\otimes {\cal A},*)$ by $\sigma$ projection. By the construction, all resulting theories are regarded as some background gauge fixed theories of the universal ``gauge theory". This suggests that the combination of a background field and a noncommutative field strength, which appears in Dirac-Born-Infeld action, has some universal meaning. 
As an application, we gave a geometrical interpretation of a `Seiberg-Witten map' in general backgrounds: It is regarded as a ``gauge transformation" and its gauge equivalence relation is a part of automorphisms of the Weyl bundle.\\

In \S\ref{sec:gfiw} we did not write an action of the ``gauge theory", because we do not define the trace of the algebra. By defining suitable trace, it may be written as $Tr[({\hat F}_A)^n]=Tr[Pf{\hat F}_A]$. Actions of noncommutative gauge theories might be given by this action through the ``gauge" fixing procedure. \\

In this paper, we mainly treated {\it fiberwise} automorphisms of $W(L,{\cal A})\otimes {\scriptstyle \bigwedge}$ because we would like concentrate on gauge theories. In this case, the geometry of the original base manifold $M$ is seen to be deformed only by ${\cal O}(\h)$. If we consider whole automorphisms, which include diffeomorphisms, we might obtain noncommutative gauge and gravity theory. In fact, in eq.(\ref{eqn:gengtry}), we treated a central term in noncommutative gauge transformation as rather trivial, but if a diffeomorphism is included, we cannot ignore it. Therefore that central term may generate a diffeomorphism, or a local transformation of 1-form basis ${\tilde \theta}^I$. In that case, our fixed basis ${\tilde \theta}^I$ (or noncommutative {\it forms}) should be treated covariantly.\\

Optimistically, our noncommutative gauge theories may be applicable to $N$-coincident D$(2n)$-branes in nonconstant B-field (NSNS 2-form) and curved backgrounds in string theory. In this case, the deformation parameter $\h$ may be taken as $\alpha'$ so that the deformation is a sort of stringy correction. By construction, $U(N)$ adjoint matter, which corresponds to the Higgs field on $N$-coincident D-branes, are introduced naturally. Fermion might also be included. If we include diffeomorphisms, our field strength (\ref{eqn:fagamm}) could be regarded as that of noncommutative Dirac-Born-Infeld action (\S\ref{construct}), and has universal meaning. 

\section*{Acknowledgements}
We would like to thank N.~Ikeda, S.~Imai, T.~Kawano and K.~Okuyama for valuable discussions and comments. 
T.~A.\ and I.~K.\ are supported in part by the Grant-in-Aid (\#04319) and (\#9858), respectively, from the Ministry of Education, Science, Sports and Culture.

\appendix
\section{Example of $*$ Product
\label{sec:ex}}
We give an example of Fedosov's $*$ product explicitly.
First, note that eq.(\ref{eqn:aita}) is solved as
\begin{eqnarray}
\label{eqn:qsolve}
&&a=Q(a_0)\nonumber\\
&&=\sum_{
\begin{subarray}{c}
l\geq0,n\geq0\\
n=n_1+n_2+\cdots+n_{l+1}
\end{subarray}
}(\delta^{-1}\nabla)^{n_1}\delta^{-1}[{i\over\h}r,(\delta^{-1}\nabla)^{n_2}\delta^{-1}[{i\over\h}r,\cdots,(\delta^{-1}\nabla)^{n_l}\delta^{-1}[{i\over\h}r,(\delta^{-1}\nabla)^{n_{l+1}}a_0]\cdots]].\nonumber\\
\end{eqnarray}
For simplicity, we consider a case that $N=1$ (i.e. $U(1)$ case), ${\Gamma^i}_j=0$ and ${\theta^i}_{\mu},\omega_{ij},R_{Eij},\Omega_{1ij}$ are constants.
Moreover we take $\mu$ as quadratic in $y$:
\begin{equation}
\mu=\frac12 \h \mu_{ij}y^i y^j,\ \ \mu_{ij}={\rm constant}.
\end{equation}
Then $r$ is determined as
\begin{eqnarray}
r_s&=&\h \mu_{ij}y^i\theta^j,\nonumber\\
r_a&=&{1\over2}\left(-i\h R_E-\Omega_1+\h^2\mu\omega^{-1}\mu\right)_{ij}y^i\theta^j+\delta^{-1}\left(\h \mu_{ij}\omega^{ik}\theta^j\wedge{\partial\over\partial y^k}r_a+{1\over2}{\partial\over\partial y^i}r_a\wedge\omega^{ij}{\partial\over\partial y^j}r_a\right),\nonumber\\
\end{eqnarray}
i.e.,
\begin{equation}
\label{eqn:rapro}
r={1\over2}(2\h \mu -i\h R_E-\Omega_1)_{ij}y^i\theta^j+W_5=r^{(2)}_{ij}y^i\theta^j,
\end{equation}
where $r^{(2)}_{ij}\in W_2$ is constant, i.e., $r$ is linear in $y^i$.
Because $r$ has no $x^\mu$ dependence in this case, (\ref{eqn:qsolve}) for $a_0=x^\mu$ is given as follows:
\begin{eqnarray}
Q(x^\mu)&=&x^\mu+\sum_{l=0}^\infty\left(i\over\h\right)^l\delta^{-1}[r,\delta^{-1}[r,\cdots,\delta^{-1}[r,y^i{X_i}^\mu]\cdots]]\nonumber\\
&=&x^\mu+y^i{\left({1\over1+(\omega^{-1}r^{(2)})^T}X\right)_i}^\mu \ .
\end{eqnarray}
For a general function $f(x)$, $Q(f(x))=f(Q(x))$ in this case.\\

$*$ product between $x^\mu$ and $x^\nu$, which also defines the parameter $\vartheta^{\mu\nu}$, is given as follows:
\begin{eqnarray}
x^\mu*x^\nu&=&x^\mu x^\nu+{i\over2}\vartheta^{\mu\nu}\nonumber\\
\vartheta^{\mu\nu}&:=&-i(x^\mu*x^\nu-x^\nu*x^\mu)=
      -i\sigma([Q(x^\mu),Q(x^\nu)])\nonumber\\
      &=&-\h\left(X^T{1\over{1+ \omega^{-1}r^{(2)}}}\omega^{-1}{1\over{1+ (\omega^{-1}r^{(2)}})^T}X\right)^{\mu\nu} \ .
\end{eqnarray}\\

Using this $\vartheta^{\mu\nu}$, we can write $*$ product as follows:
\begin{equation}
f(x)*g(x)=f(x)\exp\left({i\over2}\vartheta^{\mu\nu}\overleftarrow{\partial_\mu}\overrightarrow{\partial_\nu}\right)g(x),\quad\quad f(x),g(x)\in C^{\infty}(M)[[\h]].
\end{equation}
This $*$ product coincides with the Moyal-Weyl product which is usually taken as $*$ product on $M={\mathbb R}^{2n}$.

\section{Derivation of $*$ Algebra
\label{sec:deri}}
In the case of the usual Moyal-Weyl product, partial derivative $\partial_\mu={\partial\over\partial x^\mu}$ is a derivation with respect to it
\footnote{
Here we call it a derivation of $({\tilde A},{\tilde *})$ if an operation ${\tilde \partial}:{\tilde A}\rightarrow{\tilde A}$ satisfies ${\tilde \partial}(a+b)={\tilde \partial}a+{\tilde \partial}b, \ {\tilde \partial}(a{\tilde *}b)=({\tilde \partial}a){\tilde *}b+a{\tilde *}{\tilde \partial}b,\ {\rm for}\ \forall a,b\in{\tilde A}$.
} 
because it has the form $\exp\left({i\over2}\vartheta^{\mu\nu}\overleftarrow{\partial_\mu}\overrightarrow{\partial_\nu}\right)$, where $\vartheta^{\mu\nu}$ is constant. 
However, in general, it is {\it not} a derivation of $(C^\infty(M)[[\h]]\otimes {\cal A},*)$  
\footnote{
Although $\nabla_{L\mu}$ is a derivation of $(W_D,\circ)$, it is not derivation with respect to $*$ product.
}.
We here construct derivations of $W_D$ from inner derivations of $W(L,{\cal A})$
\footnote{
On the construction here, see, for example, Appendix A of \cite{Xu}.
}.\\

For any section $K\in W(L,{\cal A})$, an operator ${\hat \partial}_K:W(L,{\cal A})\rightarrow W(L,{\cal A})$ defined by
\begin{equation}
\label{eqn:inn}
{\hat \partial}_Ka:={i\over\h}[K,a],\ \ a\in W(L,{\cal A})
\end{equation}
is a (inner) derivation of $(W(L,{\cal A}),\circ)$.
Such ${\hat \partial}_K$ is also a derivation of $(W_D,\circ)$ if and only if
\footnote{
Note that $D({\hat \partial}_Ka)=D\left({i\over\h}[K,a]\right)={i\over\h}[DK,a]+{i\over\h}[K,Da]={i\over\h}[DK,a],\ \forall a\in W_D$
}
\begin{equation}
[DK,a]=0,\ \ \forall a\in W_D,
\end{equation}
which is equivalent that $DK$ is in center \cite{Fedbk}:
\begin{equation}
\label{eqn:DKth}
DK=\Theta,\ \ \Theta\in Z\otimes {\scriptstyle \bigwedge}^1.
\end{equation}
Because $D$ is Abelian, $d\Theta=D\Theta=D^2K=0$, i.e., $\Theta$ is closed 1-form. Locally, we may write it as $\Theta=d\Phi,\ \Phi\in Z$ and (\ref{eqn:DKth}) can be rewritten $D(K-\Phi)=0$ i.e., $K-\Phi\in W_D$. Therefore, we can write (\ref{eqn:inn}) as follows
\footnote{
In \cite{Xu}, normalization condition $\sigma(K)=0$ is adopted, but here we do not impose it.
}:
\begin{eqnarray}
K&=&Q(\sigma(K)-\Phi)+\Phi,\nonumber\\
{\hat \partial}_Ka&=&{i\over\h}[Q(\sigma(K)-\Phi),a],\ \ a\in W_D.
\end{eqnarray}
This means that ${\hat \partial}_K$ is locally a inner derivation. Therefore, it also induces a derivation of $(C^\infty(M)[[\h]]\otimes {\cal A},*)$: Write $a\in W_D$ as $a=Q(a_0)$, we define $\partial_{*K}$ by
\begin{eqnarray}
{\hat \partial}_KQ(a_0)&=&{i\over\h}[Q(\sigma(K)-\Phi),Q(a_0)], \nonumber\\
\rightarrow \ \partial_{*K}a_0&:=&\sigma({\hat \partial}_K Q(a_0))={i\over\h}[\sigma(K)-\Phi,a_0]_*,
\end{eqnarray}
where we used the notation: $[a_0,b_0]_*:=a_0*b_0-b_0*a_0$. 
$\partial_{*K}$ is obviously a derivation with respect to $*$ product.\\

\end{document}